# CDEMapper: Enhancing NIH Common Data Element Normalization using Large Language Models

Yan Wang, PhD[1#], Jimin Huang, MS[1#], Huan He, PhD[1], Vincent Zhang, MS[1], Yujia Zhou, MS[1], Xubing Hao, MS[2], Pritham Ram, MS[1], Lingfei Qian, PhD[1], Qianqian Xie, PhD[1], Ruey-Ling Weng, MS[1], Fongci Lin, PhD[1], Yan Hu, MS[2], Licong Cui, PhD[2], Xiaoqian Jiang, PhD[2], Hua Xu, PhD[1*], Na Hong, PhD[1*]

[1] Department of Biomedical Informatics and Data Science, School of Medicine, Yale University, New Haven, CT, USA; [2] Mcwilliam School of Biomedical Informatics, The University of Texas Health Science Center at Houston, Houston, TX, USA

# The authors contributed equally

* co-corresponding authors: hua.xu@yale.edu; na.hong@yale.edu

## Abstract

**Objective**

Common Data Elements (CDEs) standardize data collection and sharing across studies, enhancing data interoperability and improving research reproducibility. However, implementing CDEs presents challenges due to the broad range and variety of data elements. This study aims to develop an effective and efficient mapping tool to bridge the gap between local data elements and National Institutes of Health (NIH) CDEs.

**Methods**

We propose CDEMapper, a large language model (LLM)-powered mapping tool designed to assist in mapping local data elements to NIH CDEs. CDEMapper has three core modules: (1) CDE indexing and embeddings. NIH CDEs were indexed and embedded to support semantic search; (2) CDE recommendations. The tool combines Elasticsearch (BM25 similarity methods) with state-of-the-art GPT services to recommend candidate CDEs and their permissible values; and (3) Human review. Users review and select the NIH CDEs/values that best match their data elements and value sets. We evaluate the tool's recommendation accuracy against manually annotated mapping results.

**Results**

CDEMapper offers a publicly available, LLM-powered, and intuitive user interface that consolidates essential and advanced mapping services into a streamlined pipeline. It provides a step-by-step, quality-assured mapping workflow designed with a user-centered approach. The evaluation results demonstrated that augmenting BM25 with GPT embeddings and a ranker consistently enhances CDEMapper's mapping accuracy in three different mapping settings across four evaluation datasets.

**Discussions and Conclusions**

This work opens up the potential of using LLMs to assist with CDE recommendation and human curation when aligning local data elements with NIH CDEs. Additionally, this effort enhances clinical research data interoperability and helps researchers better understand the gaps between local data elements and NIH CDEs.

# Introduction

Collecting and reusing research data for biomedical and clinical research is becoming increasingly complex due to the growing volume of data and its varying granularity, structure, and semantics. This need is particularly pronounced in long-span, multiple sites studies, for instance, in a multicenter prospective study of ankylosing spondylitis[1], which was conducted from five sites (4 US sites and 1 Australian site) with follow-up of more than a decade, and the data collection was conducted every two years using four different versions of Case Report Forms (CRFs). The CRFs comprised several forms and data elements, some of the items in some forms were changed over time which made the data harmonization quite challenging. Similar situations occurred in many large-scale and long-span biomedical or clinical research[2-4]. Therefore, there is a pressing need to address this growing demand for consistent research data collection and reuse. However, a crucial step toward realizing this vision is to build a data infrastructure based on systematic and standardized frameworks.

The proposal of Common Data Element (CDE) encourages standardized research data collection and sharing, consistently representing data elements and values[5], thus enhancing data interoperability, and facilitating the secondary use of research data. CDEs are defined as "standardized, precisely defined questions paired with a set of specific allowable responses, used systematically across different sites, studies, or clinical trials to ensure consistent data collection"[6]. Extensive efforts were devoted by the National Institutes of Health (NIH) and other consortiums[7-13] to develop CDEs and promote CDE adoption in biomedical and clinical research. The NIH CDE Repository was created by the NIH in 2015[14], guided by the FAIR principles (Findable, Accessible, Interoperable, and Reusable)[15]. It is designed to support structured, both human and machine-readable definitions of data elements that have been recommended or required by NIH Institutes and Centers and other organizations to use in research[16]. In addition, NIH released the National Institutes of Health Data Management and Sharing (DMS) policy, effective January 25, 2023, to encourage data sharing to the extent that it is possible. This new policy requires the submission of a DMS plan with all NIH-funded research submissions, with an expectation of compliance and adherence to the plan (with allowances made for updates) throughout the lifecycle of funded projects[17]. With the DMS policy effective, the NIH recommends using CDEs, from which standard data collection forms and data elements can be developed and shared across research studies. Investigators can standardize data collection, combine sets of CDEs consisting of individual or more complex questionnaires across multiple studies, and compare results across sites[18], which in the long run enables research reproducibility[19].

However, NIH CDEs have not achieved widespread adoption yet. Several factors contribute to the limited adoption of existing CDEs. On the one hand, policy and data culture play a significant role in adopting CDEs. For example, some existing CDEs are developed for specific projects or collaborations, with limited policies and dissemination plans for continuous reuse after project completion, although there are some guiding principles to encourage reuse. Additionally, researchers may lack awareness of standardized practices, particularly in exploratory research initiated by individual investigators or research labs. On the other hand, from a technical perspective, mapping various data elements to CDEs is challenging due to the wide range of variability and the complexity involved—particularly when dealing with data elements defined at different levels of granularity. For instance, the existence of similar CDEs with insufficient context makes researchers confused when determining which is the best match; a value set using particular terminology that differs from the terminology of the source value set also complicates the selection process[14], and sometimes, a complex harmonization situation leads to a tendency to create new CDEs rather than harmonize existing ones. These various factors collectively pose challenges that hinder CDE implementation and adoption.

The traditional approach to harmonizing different data elements is conducting manual transformation from source to target data elements, which is time-consuming and needs extensive domain knowledge. An efficient mapping tool can aid in this non-trivial manual transformation and/or mapping task, reducing human effort in the process. Related work about developing mapping tools and algorithms currently exists, including (1) IMI-CDE[20], an interactive tool development for collaborative mapping of study variables to CDEs; (2) The GAAIN Entity Mapper[21], an intelligent software assistant for automated data mapping across different datasets or from a dataset to a common data model; (3) eleMap[8], a tool developed by the eMERGE program[22] to facilitate harmonization of phenotype data dictionaries and CDEs to terminological and metadata healthcare standards for interoperable representation of phenotype data; (4) D2Refine platform[23] which provides a web-based environment to create clinical research study data dictionaries and enables standardization and harmonization of its variable definitions with controlled terminology resources; and (5) comparison of different algorithms for mapping of Alzheimer's disease related data elements to the NIH CDEs[24], etc. Nevertheless, existing tools and algorithms have not been widely adopted, partially due to suboptimal performance on the mapping algorithms, limited to project-internal users only, or some studies have focused on algorithm comparisons but have not developed a practical tool. Recently, Large Language Models (LLMs) demonstrated exceptional capability and have been explored for application in biomedical data standardization tasks[25-28]. This led to our motivation to use LLMs to address these mapping challenges.

Therefore, in this study, to provide a practical tool and minimize human efforts during the mapping process, we developed CDEMapper, a publicly available, user-friendly, and LLM-powered mapping tool, for biomedical and clinical researchers to map their local data elements to the NIH CDEs. Our tool not only integrated state-of-the-art LLM technologies into a streamlined pipeline but also adopted a modular-based architecture to allow quick updates of software components and support future versions of NIH CDEs.

## Methods

### CDEMapper architecture and core modules

CDEMapper is a web-based online tool designed to assist researchers in mapping their local data elements to NIH CDEs. It indexed multiple levels of NIH CDE information to support semantic search and integrated LLM-backed multiple enhancement components. The tool is built on three core modules: CDE indexing, CDE candidate ranking, and interactive mapping review.

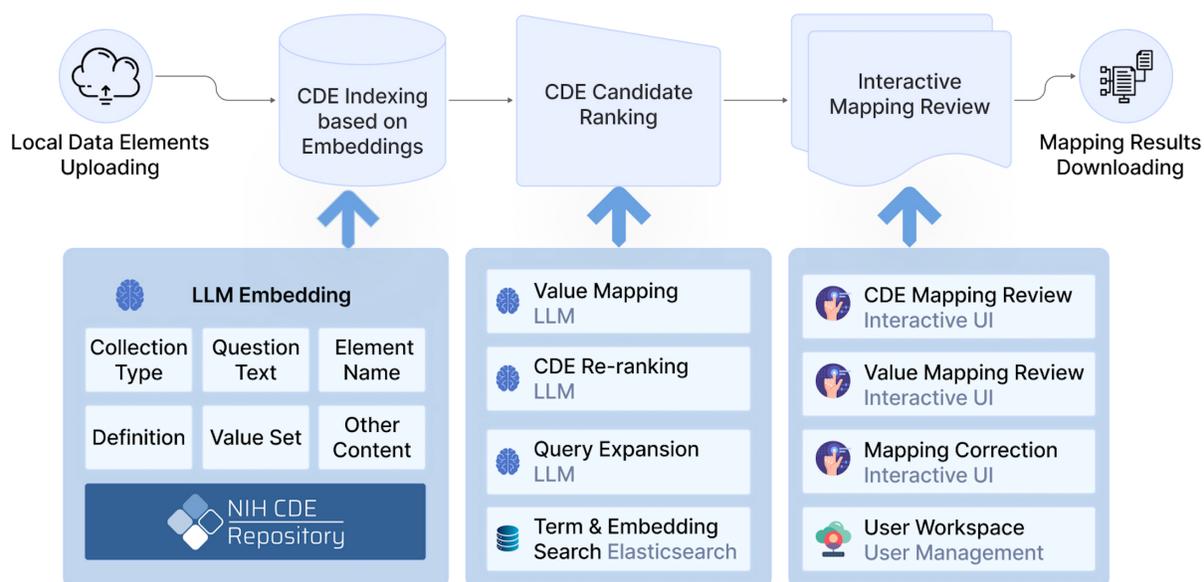

**Figure 1**. Overview of CDEMapper architecture

***CDE indexing based on embeddings.*** We downloaded and indexed all 23,328 CDEs from 19 Collections of the NIH CDE Repository as of October 21, 2024. Firstly, this module preprocessed various CDE components, including collection types, element names, question texts, definitions, value sets, and other related semantic information. Building on top of this data foundation, we leverage a GPT-based embedding model to generate semantic embeddings that integrate key CDE components such as element names, definitions, and value sets. The embeddings are created by feeding the processed CDE content into the GPT model, where its transformer architecture captures both syntactic and semantic relationships. The model computes high-dimensional vector representations by encoding contextual nuances and semantic patterns, while attention mechanisms weigh the importance of different components. By combining these elements, the embeddings effectively address lexical variation and reduce ambiguity, capturing nuanced relationships and thematic similarities across CDE components. These embeddings are further indexed in Elasticsearch with the BM25 similarity algorithm to enhance the accessibility of relevant CDEs, providing a robust data indexing for downstream mapping tasks.

***LLM-powered candidate ranking and value mapping.*** This module integrates multiple advanced services powered by GPT-4o[29] to automatically recommend candidate CDEs and values. (1) For each source data element, our tool employs a hybrid retrieval method combining a term-based ranking model (BM25 algorithm) with an LLM-based embedding search for CDE candidate recommendations. Furthermore, we developed tailored prompts for both the query expansion and CDE re-ranking modules, leveraging the capabilities of GPT-4o to enhance recommendation accuracy. For the query expansion function, the system refines and rephrases search terms, descriptions, and associated metadata to optimize their compatibility with Elasticsearch's BM25 algorithm. This process ensures that the expanded queries capture semantic equivalences and contextual nuances common in medical terminologies, such as variations between terms like "High Blood Pressure" and "Hypertension." By doing so, the system improves the recall of relevant

CDEs for the initial query results. (2) For the CDE re-ranking function, GPT-4o is employed to re-evaluate and re-prioritize the candidate CDEs recommended by Elasticsearch. This re-ranking considers both the semantic similarity between the source data element and the CDE descriptions, as well as contextual alignment with the specific research or clinical scenario. As a result, the most relevant and contextually appropriate CDEs are highlighted and ranked at the top, enhancing the precision of the recommendations. (3) To address cases where both the source and target data elements include permissible values, a GPT-powered value mapping component has been integrated into our tool. This component supports mapping source values or codes to standardized values defined in the corresponding target CDE. The prompts designed to invoke GPT 4o services are illustrated in Figure 2.

**Query expanding**

Instruction: Your task is to enhance and rephrase the provided search terms and descriptions to optimize them for querying in Elasticsearch using BM25, ensuring they account for semantic equivalences in medical terminology. Identify and include synonyms, abbreviations, and related terms for each medical term or description, such as using "myocardial infarction" for "heart attack." Rephrase the descriptions to include contextually relevant information while maintaining coherence and medical accuracy, ensuring that the expanded queries remain aligned with the original intent and meaning. For example, expand "diabetes complications" to include terms like "diabetic complications," "hyperglycemia effects," and "long-term effects of diabetes." Similarly, rephrase descriptions to include detailed aspects, such as expanding "Symptoms and management of type 2 diabetes" to "Symptoms and management of type 2 diabetes, including hyperglycemia control, insulin resistance treatment, and lifestyle changes for diabetes." The goal is to create comprehensive and precise queries and descriptions that enhance search accuracy in Elasticsearch using BM25.

Input: {"term": query_term, "description": term_description}

Output: Return only the JSON dict of search string for terms and descriptions.

**CDE Re-ranking**

Instruction: Your task is to rerank the provided search results, which are based on Elasticsearch using BM25. These results need to be adjusted to account for semantic equivalences in medical terminology. Rerank the search results based on their relevance to the query, considering this special case.

Input: {"term": query_term, "description": term_description}

Search Results: [candidate1, candidate2, …, candidate10]

Output: Return only the JSON list of reranked search results.

**Value mapping**

Instruction: Your task is to identify the most closely matched concept from the provided value sets given a value name and recalculate the score for each candidates according to semantic similarity.

Input: {"value name": value_name}

Value Set: {value 1, value2, …}

Output: Return only the JSON list of top 1 matched records ordered by recalculated semantic similarity scores.

**Figure 2**. Prompts designed for query expanding, CDE re-ranking, and value mapping

*Interactive mapping review*. The interactive mapping process, combining automated recommendations with manual refinement, ensures the mappings are as effective and efficient as possible. By default, the tool returns the top 10 best-matched CDEs as automatic recommendations for each user data element. These recommendations are sourced from various NIH repositories, allowing users to configure the data collections based on their needs, such as NIH-endorsed CDEs, NEI, NINDS, and others. The mapping review process is simple and straightforward. Users can review the top 10 recommendations and select the most relevant match. The tool also offers an extended manual search feature to support another round check for those local data elements that may not have a direct match within the algorithm recommendations. This feature allows users to perform custom searches across target CDE collections. By inputting custom search terms, users can manually explore the CDE collections and map complex cases or non-standard named data elements to the most suitable CDEs. Additionally, users can access detailed information for each suggested CDE via a hyperlink to the CDE display webpage on the NIH CDE repository, enabling informed decision-making.

**User-centered mapping workflow and interface design**

CDEMapper integrates both essential and advanced services into a user-centered mapping workflow, see Figure 3, allowing users to select different mapping strategies based on their project's needs or preferences. The BM25-based search serves as the primary mapping strategy, while advanced LLM services (highlighted as blue nodes in Figure 3), such as embedding search, query expansion, and GPT re-ranking module, are available for users to activate as needed. Users can conduct value mapping through manual review or choose LLM-assisted value mapping to receive equivalent values recommended by GPT-4o. After user confirmation, the mapping results can be exported for secondary reuse. The user can choose to map the uploaded data elements one by one for thorough review or map all at once to make the results promptly available for review and download.

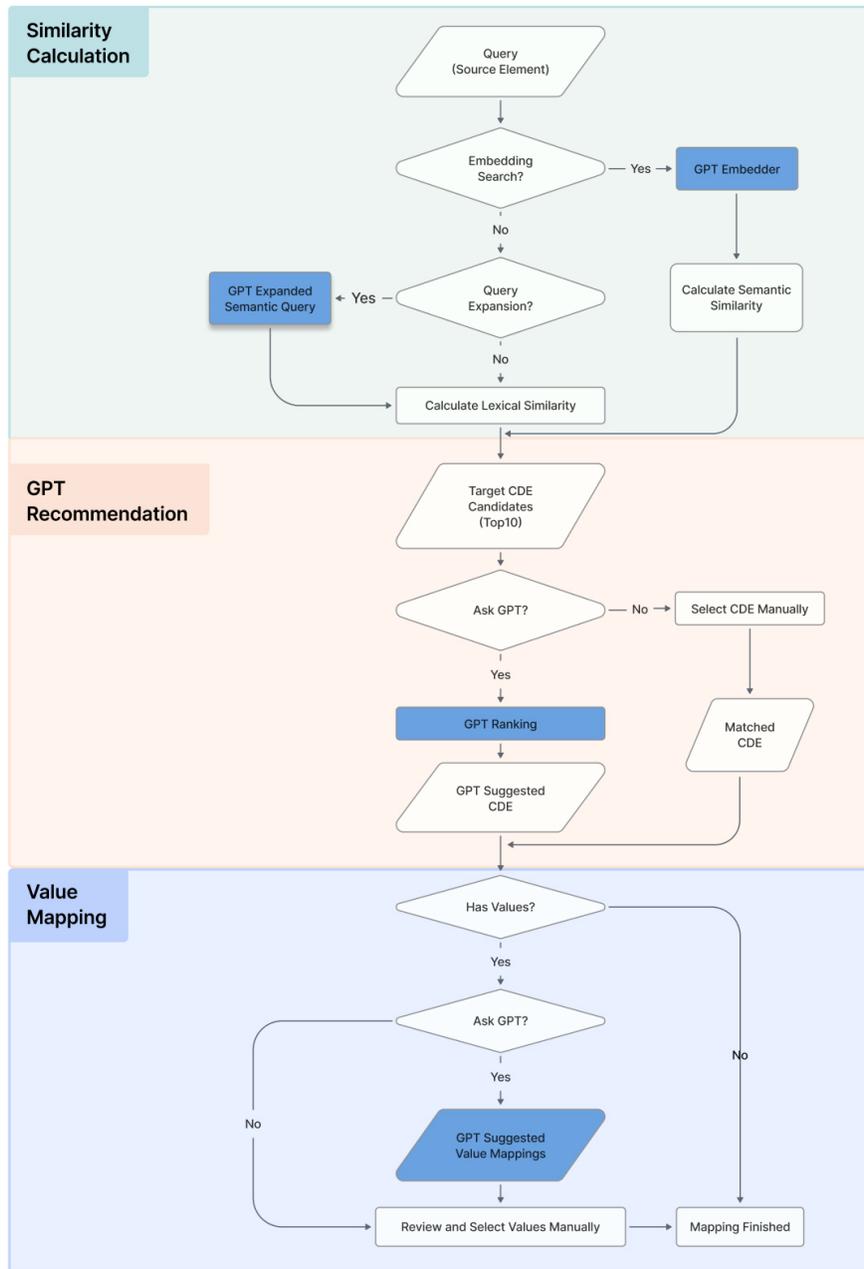

**Figure 3**. The user-centered mapping workflow

CDEMapper offers an intuitive user interface, making it easy for users from various domains to navigate and use, as shown in Figure 4. ***The Top Menu*** (1) supports the quick navigation of mapping features by intuitive buttons. ***The Source Element Panel*** (2) displays the information of source data elements, includes element names, descriptions, and permissible values, etc., and offers filtering, sorting, and pagination for enhanced navigation and control over the mapping process. ***The Target Element Panel*** (3) shows candidate CDEs, featuring the top 10 recommended results by BM25-based methods with the enhanced LLM-suggested results. The tool provides an intuitive interface for reviewing and comparing recommendations. A hyperlink is provided to navigate to the NIH CDE detailed information webpage for each candidate CDE, facilitating further user review and verification. Users can sort the data from the column headers and filter columns to work with subsets of CDEs. ***The Value Mapping Panel*** (4) supports the user in finding the most matched target values for the source value set of specific data elements. Target values can be automatically filled in by GPT recommendations, and users review the mappings.

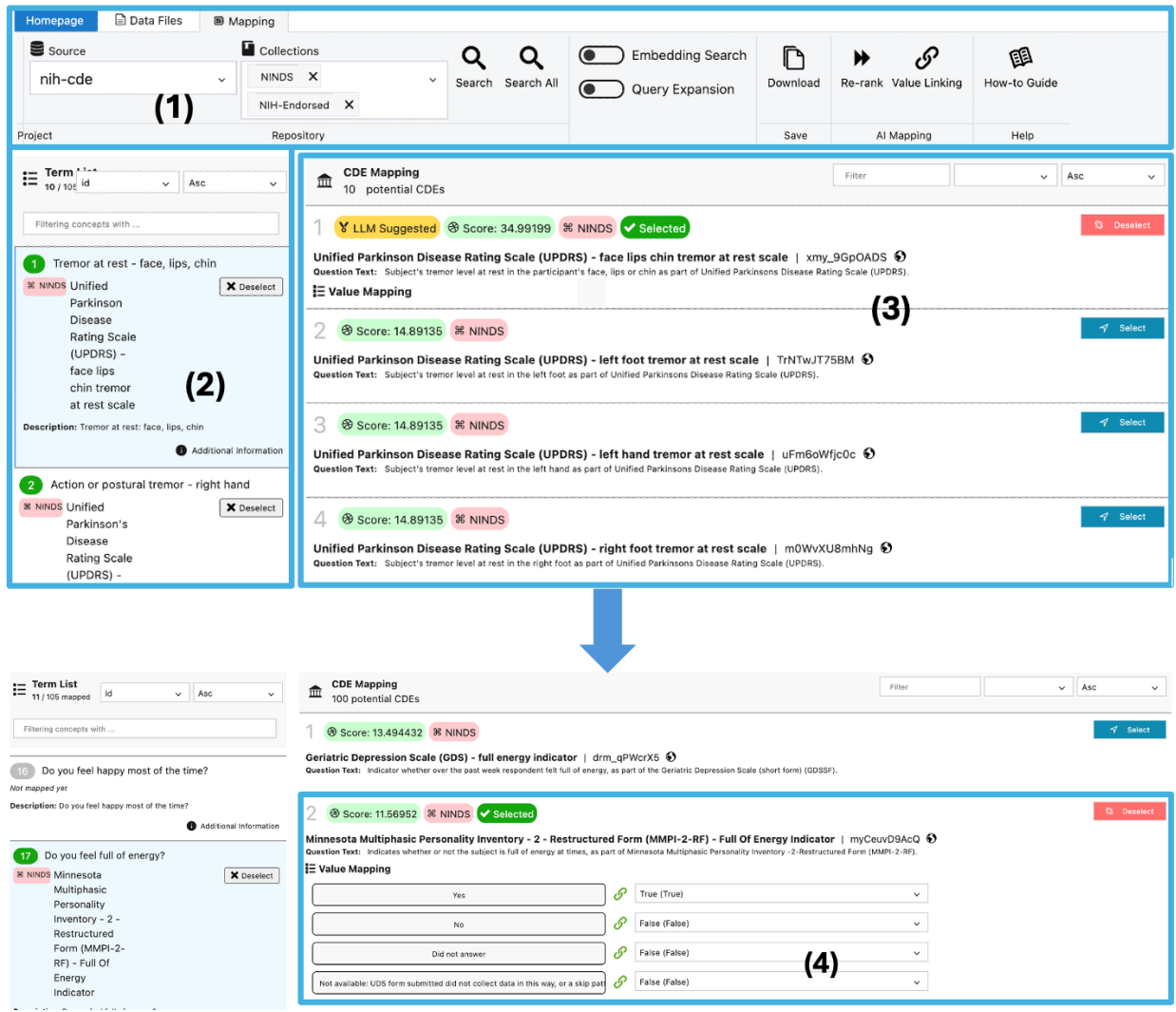

**Figure 4**. The user interface of CDEMapper, including (1) a ribbon menu for mapping actions; (2) a panel for displaying source data element awaiting mapping; (3) a panel displaying the candidate Top 10 target CDEs; and (4) an area under selected CDEs that for mapping source values to target values.

**Evaluation Methods**

To evaluate the tool's mapping capability, we collected four domains of data dictionaries or date elements, referred to as Eye (American Academy of Ophthalmology IRIS ® Registry (Intelligent Research in Sight) Analytics Data Dictionary[30]), Stroke (Human Circadian/Diurnal Biology and Stroke Common Data Elements[31]), ADRD (The National Alzheimer's Coordinating Center - Uniform Data Set (UDS)[32]), and COVID-19 (COVID-19 Therapeutic Trial Common Data Elements[33]). The four datasets contained a total of 494 data elements and have been manually mapped to different domain-related NIH CDE collections for coverage assessment and tool performance evaluation. During the manual mapping process, we started by addressing the diverse data representations and formats in the four data dictionaries through preprocessing steps such as deduplication and data formatting. Furthermore, we aimed to create all possible mappings wherever the source and target data elements are equivalent. However, not all source data elements had direct matches with NIH CDEs. If a target CDE was not found, no mapping was created for that entry. Three mapping settings were established based on the annotators' manual analysis, including "1 vs 1," "M vs 1," and "1 vs M." In "1 vs 1" mappings, one source element maps to a unique target element. In "M vs 1" mappings, multiple source elements map to a single target element. In "1 vs M" mappings, one source element maps to multiple target elements. Gold standards were manually established by two annotators. In cases of conflict, the annotators engaged in discussions to resolve discrepancies and reach consensus. After achieving annotation consensus, we finalized 264 entries with effective mapping results for results evaluation.

## Results

The mapping coverage based on manual annotation of four datasets is shown in Table 1. Eye dataset, 17 out of 40 elements can be mapped, resulting in a coverage rate of 42.50%. The stroke dataset has 48 elements, with 21 mappable elements, giving a coverage rate of 43.75%. The COVID-19 dataset contains 301 elements, of which 123 can be mapped to target CDE collections, resulting in a coverage rate of 40.86%. For the ADRD dataset, mapping coverage was not assessed in this study because the effective mappings were reused from our previous research, where the mapping coverage was reported as 9.74%[24]. After adjusting the target CDE mapping collections for this dataset, 103 out of 105 mappings were successfully reused in the ADRD dataset.

**Table 1.** The mapping coverage analysis

| Datasets | Mapped CDE Collections | Mapping examples | Numbers of mapped data elements (mapping coverage) |
|---|---|---|---|
| Eye | NIH-Endorsed CDE, NEI | Race-White --> Race (NIH-Endorsed) | 17 (42.50%) |
| Stroke | NIH-Endorsed CDE, NINDS | Imaging Modality Type --> Imaging Modality Type (NINDS) | 21 (43.75%) |
| ADRD | NIH-Endorsed CDE, NINDS | Evidence of Lewy body pathology --> Lewy body pathology indicator (NINDS) | 103 (not applicable) |
| COVID-19 | NIH-Endorsed CDE, Project 5 (COVID-19) | Ethnicity --> Ethnicity (Project 5 (COVID-19)) | 123 (40.86%) |

Table 2 presents the accuracy evaluation results for the CDEMapper under three different mapping settings ("1 vs 1," "M vs 1," and "1 vs M") for four datasets (Eye, Stroke, ADRD, and COVID-19). The models compared include the baseline BM25, BM25 combined with GPT embedding (BM25 + Emb), BM25 enhanced with GPT-4o reranking (BM25 + Rank), and a combination of BM25, GPT embedding, and GPT-4o reranking (BM25 + Emb + Rank). Acc@N refers to the accuracy measure of whether the correct mapping result is included within the top N mapping results. In addition, our -prompt for reranking focuses on recommending the top 1 result as "LLM suggested" without altering the ranks of other CDE candidates. This allows us to evaluate Acc@1 accuracy after applying the GPT-4o reranking method; however, evaluating Acc@5 and Acc@10 is not applicable after reranking, as shown in Table 2.

The Eye dataset results indicate strong performance across all methods, with almost all methods achieving high accuracies across different settings. In the "1 vs 1" and "1 vs M" settings, all methods, including BM25, BM25 + Emb, BM25 + Rank, and BM25 + Emb + Rank, achieved perfect Acc@1 of 100.00%. The "M vs 1" setting showed a slight difference, with BM25 + Emb + Rank scoring 92.31% Acc@1, consistent with the performance of the other methods. The overall performance for the Eye dataset shows that combining embeddings or using the ranker did not significantly change the performance for this dataset due to the already high baseline performance of BM25.

For the Stroke dataset, all methods performed well in the "1 vs 1" setting, with both BM25 + Rank and BM25 + Emb + Rank achieving perfect Acc@1 of 100.00%. Similarly, in the "M vs 1" scenario, BM25 + Rank and BM25 + Emb + Rank achieved 100.00% Acc@1, indicating a consistent improvement over BM25 alone, which scored 0.00%. In the "1 vs M" scenario, BM25 + Emb + Rank significantly outperformed other methods with 100.00% Acc@1 compared to BM25 and BM25 + Emb, which both scored 0.00%. Overall, for the Stroke dataset, BM25 + Emb + Rank achieved a perfect score in all three mapping settings.

In the ADRD dataset, incorporating GPT embeddings and the ranker led to considerable performance improvements. In the "1 vs 1" scenario, BM25 + Emb + Rank improved Acc@1 from 74.29% (BM25) to 81.42%. The "M vs 1" setting showed the most significant improvement: BM25 + Emb + Rank achieved 94.12% Acc@1, compared to 64.71% for the baseline BM25. In the "1 vs M" setting, BM25 + Rank and BM25 + Emb + Rank both achieved 93.75% Acc@1, indicating an improvement over the baseline BM25 (87.50%). Across all mappings, the results demonstrated the effectiveness of combining mapping, embedding, and ranking components for improving accuracy in the ADRD dataset.

In the COVID-19 dataset, the BM25 + Emb + Rank method consistently outperformed the other methods in the "1 vs 1" and "M vs 1" settings, achieving 76.19% and 52.94% accuracy at Acc@1, respectively. For "1 vs 1," the BM25 + Emb method exhibited significant improvements over BM25, with Acc@1 increasing from 61.90% to 71.43%.

However, for the "M vs 1" setting, the improvements were moderate. In the "1 vs M" scenario, BM25 + Rank and BM25 + Emb + Rank methods both achieved an Acc@1 of 82.35%, which was an improvement compared to BM25 alone (76.47%). Overall, for the COVID-19 dataset, BM25 + Emb + Rank outperformed other methods, indicating the advantage of incorporating both embeddings and a ranker for mapping tasks.

Table 2. The accuracy performance for the CDE mapping under different mapping settings.

| Datasets | Category (numbers) | Method | Acc@1 (%) | Acc@5 (%) | Acc@10 (%) |
|---|---|---|---|---|---|
| Eye | 1 vs 1 (3) | BM25 | 100.00 | 100.00 | 100.00 |
| | | BM25 + Emb | 100.00 | 100.00 | 100.00 |
| | | BM25+Rank | 100.00 | - | - |
| | | BM25+Emb+Rank | 100.00 | - | - |
| | M vs 1 (13) | BM25 | 92.31 | 100.00 | 100.00 |
| | | BM25 + Emb | 92.31 | 100.00 | 100.00 |
| | | BM25+Rank | 92.31 | - | - |
| | | BM25+Emb+Rank | 92.31 | - | - |
| | 1 vs M (1) | BM25 | 100.00 | 100.00 | 100.00 |
| | | BM25 + Emb | 100.00 | 100.00 | 100.00 |
| | | BM25+Rank | 100.00 | - | - |
| | | BM25+Emb+Rank | 100.00 | - | - |
| Stroke | 1 vs 1 (18) | BM25 | 94.44 | 94.44 | 100.00 |
| | | BM25 + Emb | 94.44 | 100.00 | 100.00 |
| | | BM25+Rank | 100.00 | - | - |
| | | BM25+Emb+Rank | 100.00 | - | - |
| | M vs 1 (2) | BM25 | 0.00 | 100.00 | 100.00 |
| | | BM25 + Emb | 0.00 | 100.00 | 100.00 |
| | | BM25+Rank | 100.00 | - | - |
| | | BM25+Emb+Rank | 100.00 | - | - |
| | 1 vs M (1) | BM25 | 0.00 | 0.00 | 0.00 |
| | | BM25 + Emb | 0.00 | 100.00 | 100.00 |
| | | BM25+Rank | 0.00 | - | - |
| | | BM25+Emb+Rank | 100.00 | - | - |
| ADRD | 1 vs 1 (70) | BM25 | 74.29 | 82.86 | 87.14 |
| | | BM25 + Emb | 74.29 | 85.71 | 88.57 |
| | | BM25+Rank | 77.14 | - | - |
| | | BM25+Emb+Rank | 81.42 | - | - |
| | M vs 1 (17) | BM25 | 64.71 | 82.35 | 88.24 |
| | | BM25 + Emb | 76.47 | 88.24 | 94.12 |
| | | BM25+Rank | 76.47 | - | - |
| | | BM25+Emb+Rank | 94.12 | - | - |
| | 1 vs M (16) | BM25 | 87.50 | 93.75 | 93.75 |
| | | BM25 + Emb | 87.50 | 93.75 | 100.00 |
| | | BM25+Rank | 93.75 | - | - |
| | | BM25+Emb+Rank | 93.75 | - | - |
| COVID-19 | 1 vs 1 (21) | BM25 | 61.90 | 71.43 | 85.71 |
| | | BM25 + Emb | 71.43 | 80.95 | 80.95 |
| | | BM25+Rank | 66.67 | - | - |
| | | BM25+Emb+Rank | 76.19 | - | - |
| | M vs 1 (85) | BM25 | 42.35 | 62.35 | 69.41 |
| | | BM25 + Emb | 47.06 | 58.82 | 68.24 |
| | | BM25+Rank | 48.24 | - | - |
| | | BM25+Emb+Rank | 52.94 | - | - |
| | 1 vs M (17) | BM25 | 76.47 | 82.35 | 82.35 |
| | | BM25 + Emb | 64.71 | 94.11 | 100.00 |
| | | BM25+Rank | 82.35 | - | - |
| | | BM25+Emb+Rank | 82.35 | - | - |

The experimental results demonstrate that augmenting BM25 with GPT embedding (BM25 + Emb) and further enhancing it with a GPT 4o (BM25 + Rank) generally improves the retrieval performance across different datasets and mapping settings. The methods with both embeddings and GPT 4o (BM25 + Emb + Rank) consistently outperformed other configurations, especially for challenging mapping scenarios like "M vs 1." The Stroke and Eye datasets showed strong performance across methods, suggesting less variability in the dataset or higher mapping efficacy of the baseline. The ADRD and COVID-19 datasets, however, benefited significantly from the enhanced methods, with noticeable gains in accuracy.

## Discussions

Our work provides valuable insights into normalizing research data elements to align with NIH CDEs and highlights the potential of leveraging LLMs to enhance research data interoperability. The evaluation results demonstrated that augmenting BM25 with GPT embeddings and a GPT ranker consistently enhances CDEMapper's mapping accuracy in three different mapping settings across four evaluation datasets.

Advantages of the CDEMapper include (1) it incorporates a large extent of CDE definitions and detailed information, such as element names, question text, definitions, value sets, and other semantic information, to build the indexing; (2) it integrates advanced GPT services for several mapping recommendation tasks in the tool. Based on performance evaluation results, the best performing BM25 method, combined with GPT embeddings and a GPT ranker, has been adopted and deployed in the tool; (3) the tool takes into account version evolvement and uses modulization management. It can support future CDE versions with minor updates of indexing and embeddings. It also supports LLM upgrades, transitions to other advanced models, or prompt refinements, ensuring the sustainability of all functions and workflows. This tool will continue to be upgraded and integrated with the latest state-of-the-art mapping algorithms. The long-term goal of this research is to develop an openly accessible tool for mapping CDEs designed to serve a wide range of research communities.

Throughout this work, we gained valuable experience and learned many lessons, including (1) overall, the results indicate that the mapping coverage is generally low. This reflects the gap from real-world research data elements to NIH CDEs; however, our evaluation results based on four domain data dictionaries are not fully representative of all research cases because each study has different variables, data dictionaries, and data elements; (2) generally speaking, the current CDEs development is immature. Some NIH CDE collections were submitted as common research data elements of a specific domain/project into the NIH CDE repository. They typically focused on different research purposes, resulting in varying levels of granularity in data elements and value set definitions. This challenged the mapping tool with unstable performance; (3) although much research emphasizes the importance of CDEs, there is a lack of studies focusing on their implementation. Developing an effective and publicly accessible mapping tool could enhance the discovery and adoption of CDEs; and (4) designing interactive workflows for mapping review in a user-friendly manner is crucial to ensure the tool's usability, efficient human-in-the-loop processes, and high-quality mappings deliverables.

Our study has limitations. (1) With the assistance of the GPT-4o ranker, our tool improved performance in some challenging cases that BM25 could not effectively identify. However, since our tool relies on a pipeline-based method to map CDE, the search results from BM25 greatly influence subsequent performance. (2) We also developed a query expansion function utilizing GPT-4o to enhance query expressions. However, the recommendation performance did not improve as expected after expanding the query with GPT-4o across all datasets except the COVID-19 dataset. This method needs to be further refined and tested. (3) In the current CDEMapper version, the uploaded file is limited to CSV format, and three main columns of semantic information from source data elements need to be loaded into the tool for semantic mapping; (4) We did not differentiate between various mapping types, such as exact match, partial match, broad match, and narrow match. In this study, some partial mappings, even if not entirely identical, were also considered valid mappings. (5) We have not conducted a performance evaluation of value mapping based on GPT recommendations due to the lack of annotated mapping datasets.

In the future, we plan to improve the following four aspects: (1) Train a CDE-specific embedder to refine our hybrid method to integrate BM25 and embeddings better, thereby improving performance in cases with significant semantic differences. (2) We plan to test additional datasets, including data element mapping and value set mapping evaluations, using our tool to gain a deeper understanding of the gap between real-world data elements and NIH CDEs, from coverage to semantics. (3) In addition, identifying truly common elements is an important task for future CDE development and harmonization, defining CDEs with high-quality and consistent semantics will benefit their long-term adoption. Therefore, we plan to expand the tool's functionality to assist domain experts in developing domain CDEs after pinpointing 'off the shelf' CDEs through our CDEMapper tool.

## Conclusions

This effort contributes to the development of algorithms and tools for mapping local data elements to NIH CDEs. The findings of this study demonstrate that the combined matching algorithm (BM25+GPT) outperforms the standalone BM25 method and can effectively facilitate the tool's mapping tasks. In addition, CDEMapper offers a publicly accessible, interactive interface built on a user-centered mapping workflow, enabling effective and efficient user engagement. In conclusion, CDEMapper will promote the reuse of NIH CDEs by providing a practical tool. The widespread adoption of CDEs, in turn, enhances interoperability across different studies and gradually ensures consistent data collection and sharing.

## Acknowledgments

This work was supported in part by grant U24LM013755 from NIH/NLM.

## Conflicts of interest

The authors have no competing interests to declare.

## Data availability

The CDEMapper tool is available online and can be accessed via the URL[1]. The evaluation datasets and code are available at the GitHub repository[2].

---

[1] https://cdemapper.clinicalnlp.org/
[2] https://github.com/BIDS-Xu-Lab/CDE-Mapping-Tool